\documentclass[aps,prd,nofootinbib,twocolumn,amsmath]{revtex4-1}
\usepackage{times,amsbsy,amsmath,amsfonts,graphicx,float}
\usepackage{color,morefloats,rotating,srcltx,slashed}
\usepackage{multirow,bm,verbatim,tabularx,bbding,threeparttable}
\usepackage{braket}
\usepackage{enumitem}
\usepackage{float}
\usepackage{xcolor}
\usepackage{mathtools}
\definecolor{dblue}{rgb}{0.00,0.00,0.75}
\usepackage[colorlinks,urlcolor=dblue,linkcolor=dblue,citecolor=dblue]{hyperref} 
\usepackage{multirow}
\allowdisplaybreaks[4]
\renewcommand{\arraystretch}{1.2}
\newcommand{\PreserveBackslash}[1]{\let\temp=\\#1\let\\=\temp}
\newcolumntype{C}[1]{>{\PreserveBackslash\centering}p{#1}}
\newcolumntype{R}[1]{>{\PreserveBackslash\raggedleft}p{#1}}
\newcolumntype{L}[1]{>{\PreserveBackslash\raggedright}p{#1}}
\usepackage{xcolor}
\newcommand{\gG}{g_iG^{II}_i}

\begin{document}

 \title{Predicted Exotic Doubly Heavy-Strange Pentaquarks}

\author{A. Feijoo}
	\email{edfeijoo@ific.uv.es}
	\affiliation{Departamento de F\'{\i}sica Te\'orica and IFIC, Centro Mixto Universidad de
	                 Valencia-CSIC Institutos de Investigaci\'on de Paterna, Aptdo.~22085, 46071 Valencia, Spain}

\author{E. Oset}
	\email{oset@ific.uv.es}
	\affiliation{Departamento de F\'{\i}sica Te\'orica and IFIC, Centro Mixto Universidad de
       		          Valencia-CSIC Institutos de Investigaci\'on de Paterna, Aptdo.~22085, 46071 Valencia, Spain}

\begin{abstract}
We predict exotic doubly heavy--strange pentaquarks with minimal quark content $u\bar d sQQ'$ ($QQ'=cc,bc,bb$) within a coupled-channel unitary framework where the interaction is derived from an extension of the local hidden gauge approach to the heavy-quark sector. We obtain a robust spectrum of manifestly exotic states; two states appear in the $u\bar d scc$ sector, three in $u\bar d scb$, and two in $u\bar d sbb$. These emerge either as bound states or resonances, along with seven additional virtual states manifested as threshold cusps. The binding mechanism is dominated by off-diagonal transitions among heavy-baryon--light-meson channels, while diagonal interactions are strongly suppressed. These results extend exotic hadron spectroscopy into the doubly heavy--strange sector and provide concrete targets for future experimental searches.
\end{abstract}

\date{\today}

\maketitle

\section{Introduction}
The experimental discovery of pentaquark candidates over the last decade has reshaped hadron spectroscopy and established exotic multiquark states as a central topic in strong-interaction physics. A major milestone was achieved by the LHCb Collaboration in 2015 with the observation of two resonant structures in the $J/\psi p$ invariant mass distribution of the $\Lambda_b\to J/\psi K^- p$ decay~\cite{LHCb:2015yax}. These states, later denoted as $P_{\psi}^N(4380)$ and $P_{\psi}^N(4450)$, cannot be accommodated within the conventional baryon spectrum because of their large masses, and necessarily require hidden charm in their minimal quark content. Interestingly, such states had already been anticipated in both molecular approaches~\cite{Wu:2010jy,Wu:2010vk,Yang:2011wz,Xiao:2013yca,Karliner:2015ina} and constituent quark models~\cite{Wang:2011rga,Yuan:2012wz}. Subsequent analyses of Run~I and Run~II data resolved the original $P_{\psi}^N(4450)$ signal into the narrow structures $P_{\psi}^N(4440)$ and $P_{\psi}^N(4457)$ and established the additional state $P_{\psi}^N(4312)$, while later measurements revealed evidence for another structure, $P_{\psi}^N(4337)$, in the $B_s^0\to J/\psi p\bar p$ decay~\cite{LHCb:2021chn}. These observations triggered extensive theoretical activity, leading to interpretations based on hadronic molecules~\cite{Chen:2015loa,He:2015cea,Liu:2019tjn,Du:2021fmf}, compact multiquark configurations~\cite{Chen:2015moa,Wang:2015epa,Wang:2019got,Ortega:2016syt,Park:2017jbn,Weng:2019ynv,Zhu:2019iwm,Deng:2022vkv}, triangle singularities~\cite{Guo:2015umn,Liu:2015fea,Mikhasenko:2015vca}, and cusp effects~\cite{Nakamura:2021dix}.

The existence of strange partners of the hidden-charm pentaquarks naturally emerged as the next step in the exploration of exotic baryons. In fact, several theoretical works had already predicted such states using unitarized extended hidden-gauge approaches and $SU(4)$ symmetry schemes~\cite{Wu:2010vk,Hofmann:2005sw,Wang:2019nvm,Xiao:2019gjd}, as well as diquark--diquark--antiquark models~\cite{Anisovich:2015zqa,Wang:2015wsa}. Motivated by these expectations, a number of studies proposed the analysis of the $J/\psi\Lambda$ invariant mass distribution in $\Lambda_b$ and $\Xi_b$ decays as a promising strategy to search for hidden-charm strange pentaquarks~\cite{Feijoo:2015kts,Lu:2016roh,Chen:2015sxa,Shen:2020gpw}. Experimental confirmation eventually came from the analyses of the $\Xi_b^-\to J/\psi\Lambda K^-$~\cite{LHCb:2020jpq} and $B^-\to J/\psi\Lambda\bar p$~\cite{LHCb:2022ogu} decays, which provided evidence for the $P_{\psi s}^\Lambda(4459)$ and $P_{\psi s}^\Lambda(4338)$ structures. Their observation stimulated renewed interest in the theoretical description of exotic hadrons and motivated both comprehensive reviews and new theoretical developments~\cite{Chen:2020uif,Chen:2020opr,Liu:2020hcv,Feijoo:2022rxf,Karliner:2022erb,Wang:2022mxy,Yan:2022wuz,Ozdem:2022kei,Ortega:2022uyu,Wang:2022tib,Peng:2020hql,Xiao:2021rgp,Zhu:2021lhd,Burns:2022uha,Nakamura:2022gtu,Meng:2022wgl,Wu:2024lud}. Within this context, states with higher strangeness become particularly compelling. Hidden-charm pentaquarks with double strangeness have already been predicted in several frameworks~\cite{Ortega:2022uyu,Wang:2020bjt,Ferretti:2020ewe,Azizi:2021pbh,Marse-Valera:2022khy,Roca:2024nsi,Song:2024yli}, although no experimental evidence has yet been reported. In addition, the transition $P_{csss}(\bar P_{csss})\to J/\psi\,\Omega^-$ associated with a possible triple-strange pentaquark configuration was investigated in Ref.~\cite{Azizi:2022qll} using QCD sum rules.

In contrast to the rapidly developing hidden-charm sector, manifestly exotic pentaquarks of the type $\ket{q_1\bar q_2 Q_1Q_2Q_3}$ with $q_1=u,d$, $q_2=d,u$ and $Q_i=s,b,c$ are still poorly studied. Interestingly, in Ref.~\cite{Gamermann:2011mq}, an $SU(6)$ extension of the meson-baryon chiral Lagrangian was implemented in a coupled-channel unitary framework, leading to the prediction of several exotic configurations associated with attractive multiplets of the model in addition to nine $\Omega^\ast$ resonances. More recently, the possible dynamical generation of a triply strange pentaquark $P_{sss}$ was investigated within a next-to-leading-order unitarized chiral framework~\cite{Feijoo:2024qgq}, where the NLO contributions were shown to provide the additional attraction required to generate the state. Complementary QCD sum-rule studies have also predicted exotic pentaquark structures with strangeness and double-heavy content~\cite{Yang:2024okq}. Related configurations were previously investigated in Ref.~\cite{Wang:2020bjt} within an effective Lagrangian framework incorporating heavy-quark, chiral, and local hidden-gauge symmetries.

In this work, we investigate the possible existence of molecular pentaquark states with minimal quark content $\ket{u\bar{d}sQQ'}$, where $QQ'=bb,cc,$ or $bc$, within a coupled-channel unitarized effective field theory (EFT) framework.  These systems carry isospin $I=1$ and are manifestly exotic, since their quantum numbers cannot be realized in conventional baryons. We show that the coupled-channel dynamics among the relevant heavy-hadron configurations provides sufficient attraction to dynamically generate such states. Our results extend the study of exotic multiquark spectroscopy into a largely unexplored sector and provide concrete targets for future searches for doubly heavy pentaquarks in high-energy experiments. \\

\section{Formalism}
As we shall see, the formation of the exotic states under consideration is driven by the presence of coupled channels. The states we study are of the type $\ket{u\bar{d}scc}$, $\ket{u\bar{d}scb}$, and $\ket{u\bar{d}sbb}$. All of them have isospin $I=1$, a quantum number that cannot be realized within conventional three-quark configurations containing $b$, $c$, or $s$ quarks alone. Moreover, these are genuine pentaquark states, since strong and electromagnetic interactions conserve flavor, and there is no $q\bar q$ pair with the same flavor that could annihilate. As a consequence, these states can only decay into other hadron systems containing at least five quarks, which suggests a certain degree of stability.
\vspace{-0.4cm}
\begin{table}[H]
\centering
\caption{Coupled channels considered and threshold masses (in MeV) for the different sectors with the corresponding quark content $u\bar d sQ_1Q_2$. The hadron masses are taken from Table~XVII of Ref.~\cite{Wang:2022aga}; see text for details.}
\label{Table_QQs}
\setlength{\tabcolsep}{8pt}
\begin{tabular}{lcccc}
\hline \hline
& \multicolumn{4}{c}{$u\bar d scc$ quark content} \\
\hline
\multirow{2}{*}{$PB(\frac{1}{2}^+)$}
& $\Omega^+_{cc}\pi^+$ & $\Xi^{++}_{cc}\bar{K}^0$  & $\Xi^+_c D^+$ & $\Xi'^+_c D^+$ \\
& $3854$ & $4115$  & $4338$ & $4448$ \\
\hline
\multirow{2}{*}{$PB(\frac{3}{2}^+)$}
 & $\Omega^{*+}_{cc}\pi^+$ & $\Xi^{*++}_{cc}\bar{K}^0$ & $\Xi^{*+}_c D^+$ & \\
 & $3911$ & $4168$ & $4516$ & \\
\hline
\multirow{2}{*}{$VB(\frac{1}{2}^+)$}
& $\Omega^+_{cc}\rho^+$ & $\Xi^{++}_{cc}\bar{K}^{*0}$ & $\Xi^+_c D^{*+}$ & $\Xi'^+_c D^{*+}$ \\
& $4485$ & $4512$ & $4478$ & $4588$ \\
\hline
\multirow{2}{*}{$VB(\frac{3}{2}^+)$}
& $\Omega^{*+}_{cc}\rho^+$ & $\Xi^{*++}_{cc}\bar{K}^{*0}$  & $\Xi^{*+}_c D^{*+}$ & \\
& $4542$ & $4565$ & $4656$ & \\
\hline \hline
& \multicolumn{4}{c}{$u\bar d sbb$ quark content} \\
\hline
 \multirow{2}{*}{$PB({\frac{1}{2}}^+)$} 
& $\Omega^-_{bb}\pi^+$ & $\Xi^0_{bb}\bar{K}^0$ & $\Xi^0_b \bar{B}^0$ & $\Xi'^0_b \bar{B}^0$ \\
    & $10369$  & $10631$ & $11076$ 	& $11214$       	 \\
     \hline
  \multirow{2}{*}{$PB({\frac{3}{2}}^+)$} 
 & $\Omega^{*-}_{bb}\pi^+$ & $\Xi^{*0}_{bb}\bar{K}^0$  & $\Xi^{*0}_b \bar{B}^0$\\
            ~     & $10397$     & $10662$ 	& $11231$		  \\
\hline
  \multirow{2}{*}{$VB({\frac{1}{2}}^+)$}
& $\Omega^-_{bb}\rho^+$ &$\Xi^0_{bb}\bar{K}^{*0}$ & $\Xi^0_b\bar{B}^{*0}$ & $\Xi'^0_b \bar{B}^{*0}$ \\
           & $11000$       & $11028$ & $11122$ 	& $11260$ 	\\
\hline
  \multirow{2}{*}{$VB({\frac{3}{2}}^+)$}
  & $\Omega^{*-}_{bb}\rho^+$  & $\Xi^{*0}_{bb}\bar{K}^{*0}$  & $\Xi^{*0}_b \bar{B}^{*0}$ \\
           & $11028$      & $11059$ 	& $11277$ 	 		  \\ 
\hline \hline
& \multicolumn{4}{c}{$u\bar d scb$ quark content} \\
\hline
  \multirow{2}{*}{$PB({\frac{1}{2}}^+)$} 
   & $\Omega^0_{bc}\pi^+$  & $\Xi^+_{bc}\bar{K}^0$   & $\Xi^0_b D^+$ & $\Xi^+_c \bar{B}^0$     \\
            ~  & $7150$        &  $7415$ 	& $7667$ 	& $ 7747$       	 \\
\hline
  \multirow{2}{*}{$PB'({\frac{1}{2}}^+)$} 
    & $\Omega'^0_{bc}\pi^+$  & $\Xi'^+_{bc}\bar{K}^0$   & $\Xi'^0_b D^+$ & $\Xi'^+_c \bar{B}^0$     \\
            ~  & $7186$        &  $7441$  	& $7805$ 	& $7857$   	 \\
\hline 
  \multirow{2}{*}{$PB({\frac{3}{2}}^+)$}
        & $\Omega^{*0}_{bc}\pi^+$  &$\Xi^{*+}_{bc}\bar{K}^0$   & $\Xi^{*0}_b D^+$ & $\Xi^{*+}_c \bar{B}^0$   \\
           & $7205$    &  $7466$ 		& $7822$ 	& $7925$	  \\
\hline
  \multirow{2}{*}{$VB({\frac{1}{2}}^+)$}
        & $\Omega^0_{bc}\rho^+$   &$\Xi^+_{bc}\bar{K}^{*0}$   & $\Xi^0_b D^{*+}$ & $\Xi^+_c \bar{B}^{*0}$  \\
            & $7781$    &  $7812$  	& $7807$ 	& $7793$	 	  \\
\hline
  \multirow{2}{*}{$VB'({\frac{1}{2}}^+)$}
& $\Omega'^0_{bc}\rho^+$ &$\Xi'^+_{bc}\bar{K}^{*0}$ & $\Xi'^0_b D^{*+}$ & $\Xi'^+_c \bar{B}^{*0}$  \\
           & $7817$        &  $7838$ 	 	& $7945$ 	& $7903$\\
\hline
  \multirow{2}{*}{$VB({\frac{3}{2}}^+)$}
   & $\Omega^{*0}_{bc}\rho^+$  &$\Xi^{*+}_{bc}\bar{K}^{*0}$  & $\Xi^{*0}_b D^{*+}$& $\Xi^{*+}_c\bar{B}^{*0}$  \\
           & $7836$        & $7863$ 	& $7962$ 	& $7971$	     \\           
\hline
\hline
\end{tabular}
\end{table}

Inspecting the possible meson--baryon configurations compatible with the quark contents above, we find the constituent combinations displayed in Table~\ref{Table_QQs}, classified according to their meson--baryon structure. In addition to the flavour-content classification, the coupled channels considered can be organized into four sectors, namely $PB(\frac{1}{2}^+)$, $PB(\frac{3}{2}^+)$, $VB(\frac{1}{2}^+)$, and $VB(\frac{3}{2}^+)$, plus the $PB'(\frac{1}{2}^+)$ and $VB'(\frac{1}{2}^+)$ sectors in the $u\bar d scb$ case, where the primed baryons $\Omega'^0_{bc}$, $\Xi'^{+}_{bc}$, $\Xi'^{+}_{c}$ and $\Xi'^{0}_{b}$ appear. Here, $P$ denotes a pseudoscalar meson, $V$ a vector meson, $B(\frac{1}{2}^+)$ the ground-state baryons with $J^P=\frac{1}{2}^+$, and $B(\frac{3}{2}^+)$ the ground-state baryons with $J^P=\frac{3}{2}^+$. In the present work, these sectors are treated independently and no mixing among them is considered, as is done, and justified, in similar studies \cite{Debastiani:2017ewu,Yu:2018yxl,Wang:2022aga,Song:2025tha}. The masses of the hadrons entering the coupled channels in Table~\ref{Table_QQs} are taken from Table~XVII of Ref.~\cite{Wang:2022aga}. In that work, the masses of the hadrons listed in the Review of Particle Physics~\cite{ParticleDataGroup:2026mpi} are used directly, while for those not listed there, the values are taken from Table~V of Ref.~\cite{Zhou:2018bkn}, which collects predictions from several theoretical models. To ensure consistency, we take the masses of all doubly bottom baryons, $\Omega^{(*)-}_{bb}$ and $\Xi^{(*)0}_{bb}$, from a single framework, namely the relativistic quark model (RQM) of Ref.~\cite{Lu:2017meb}, instead of mixing values from different models.

\begin{figure}[H]   
  \centering
  \includegraphics[width=4.5cm]{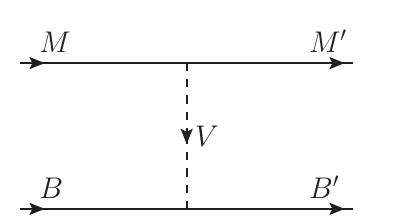}
  \caption{Diagrammatic representation of the interaction $MB\to M^\prime B^\prime$ mediated by vector-meson ($V$) exchange. $M$ ($M^\prime$) and $B$ ($B^\prime$) denote the initial (final) meson and baryon states, respectively.}
  \label{fig1}
\end{figure}

In order to construct the interaction among the coupled channels, we closely follow the formalism developed in Ref.~\cite{Wang:2022aga}, devoted to the study of the $\Omega_{cc}$, $\Omega_{bb}$, and $\Omega_{cb}$ states. Although those systems are not exotic, unlike the ones considered here, the underlying interaction mechanism is the same. We employ the extension of the local hidden gauge approach to the heavy-quark sector~\cite{Bando:1984ej,bando1988nonlinear,Harada:2003jx,Meissner:1987ge,Nagahiro:2008cv}, where the interaction is generated through vector-meson exchange between the mesons and baryons participating in the coupled channels, as illustrated in Fig.~\ref{fig1}.

The upper vertex ($VMM^\prime$) in the diagram of Fig.~\ref{fig1} is given by the lagrangians 
\begin{align}
    \mathcal{L}_{\mathrm{VPP}} &= -i g\left\langle\left[P, \partial_{\mu} P\right] V^{\mu}\right\rangle, \label{eq-1}\\
    \mathcal{L}_{\mathrm{VVV}} &=  i g\left\langle\left(V^{\mu} \partial_{\nu} V_{\mu}-\partial_{\nu} V^{\mu}
                                     V_{\mu} \right) V^{\nu}\right\rangle, \label{eq-2}    
\end{align}
where the coupling is defined as $g=\frac{m_V}{2f_\pi}$, with $m_V=800$ MeV, and the pion decay constant $f_\pi=93$ MeV. The fields $P$ or $V$ appearing in Eqs.~\eqref{eq-1} and \eqref{eq-2} are the $q_i\bar{q}_j$ matrices with $q_{i,j}=u,d,s,c,b$, expressed in terms of the corresponding meson fields (see Eqs.~(5-8) in Ref.~\cite{Wang:2022aga}).
\vspace{-0.4cm}

In Eq.~\eqref{eq-2}, when working close to threshold, the external vectors have a time component of the polarization vector satisfying $\epsilon^0\sim 0$. If $V^\nu$ corresponds to an external vector meson, then $\nu=1,2,3$, and the associated momentum $p_\nu$ from $\partial_\nu$ becomes a three momentum, which is negligible near threshold. Consequently, $V^\nu$ must correspond to the exchanged vector, as in the case of Eq.~\eqref{eq-1}. The only difference between the two interactions is then the factor $\epsilon^\mu\epsilon'_\mu=-\vec{\epsilon}\,\vec{\epsilon}'$ associated with the external vector mesons. This factor does not affect the analytical structure of the amplitudes, rendering Eqs.~\eqref{eq-1} and \eqref{eq-2} formally equivalent up to the additional polarization factor, $\vec{\epsilon}\,\vec{\epsilon}'$, in the vector-meson case.

The lower vertex in Fig.~\ref{fig1} is most conveniently evaluated following the formalism of Ref.~\cite{Debastiani:2017ewu}, where the interaction is written in terms of the quark wave functions of the exchanged vector meson and the baryons,
\begin{equation}
\label{VBB_vertex}
    \mathcal{L}_{\mathrm{VBB'}}=\bra{B'} g q_V \bar q_V \gamma^{\nu'} V_{\nu'}\ket{B}
\end{equation}
with $q_V \bar q_V$ denoting the quark content of the exchanged vector meson, while $B$ and $B'$ stand for the baryon wave functions in terms of quarks. In the approximation where the baryon three-momenta are neglected compared to the vector-meson mass, the operator $\gamma^{\nu'}$ effectively reduces to $\gamma^0\sim 1$, rendering the interaction spin independent.

For the baryon wave functions we do not rely on $SU(4)$ or $SU(5)$ symmetry. Instead, following Refs.~\cite{Capstick:1986ter,Roberts:2007ni}, the heaviest quark is singled out and the symmetry is imposed only on the lighter quark sector. A complete compilation of the wave functions employed here can be found in Table~IV of Ref.~\cite{Wang:2022aga}. An interesting feature emerges in the $PB(\frac12^+)$ sector with $u\bar d s c c$ content shown in Table~\ref{Table_QQs}, where the channels $\Omega_{cc}^+\pi^+$ and ${\Xi'}_c^+ D^+$ couple to each other (analogously to $\Omega_{bb}^-\pi^+$ and ${\Xi'}_b^+ \bar B^0$ in the $u\bar d s b b$ sector). This is a consequence of the fact that the $\Omega_{cc}^+$ state involves the spin wave function $\chi_{MS}(12)$, whereas ${\Xi'}_c^+$ contains $\chi_{MA}(23)$, which are not orthogonal, as reflected in
\begin{align}
  \langle \chi_{MS}(12)| \chi_{MS}(23) \rangle &= -\frac{1}{2} \, , \\
  \langle \chi_{MS}(12)| \chi_{MA}(23) \rangle & =-\frac{\sqrt{3}}{2}.
\end{align}
In contrast, the $\Omega_{bc}^0$ and $\Omega_{bc}^{\prime 0}$ states are described by the spin-flavor wave functions 
\begin{align}
    \ket{\Omega^0_{bc}} &= \frac{1}{\sqrt{2}} b(cs-sc)\ket{\chi_{MA}(23)} \, , \\
    \ket{\Omega'^0_{bc}} &= \frac{1}{\sqrt{2}} b(cs+sc)\ket{\chi_{MS}(23)} \, ,
\end{align}
which are orthogonal not only to each other but also to the remaining baryon states possessing a different symmetry structure in the light-quark pair. Consequently, these states give rise to two independent non-interacting sectors, denoted in Table~\ref{Table_QQs} as the $PB(\frac12^+)$ and $PB'(\frac12^+)$ sectors for the $u\bar d s c b$ quark content, and analogously for the corresponding $VB(\frac12^+)$ and $VB'(\frac12^+)$ sectors.

Finally, independently of the sector considered, the interaction kernel describing the transition from an initial channel $i$ to a final channel $j$, derived from the topology depicted in Fig.~\ref{fig1}, can always be written as
\begin{eqnarray}
  \label{eq:def_Vij}
     V_{ij}= -\frac{1}{4f_\pi^2}(p^0+p'^{0})C_{ij} \, ,
\end{eqnarray}
where $p^0$, $p'^{0}$ are the energies of the initial and final mesons, respectively, in the meson-baryon rest frame. The coefficients $C_{ij}$ are straightforwardly evaluated and are collected in the corresponding tables of Appendix~\ref{appendix}. The form of Eq.~\eqref{eq:def_Vij} follows from the fact that the derivatives $\partial_\mu$ and $\partial_\nu$ appearing in Eqs.~\eqref{eq-1} and \eqref{eq-2} effectively reduce to their temporal components, $\partial_0$, in the limit of small three-momenta. 

At first sight, the use of Eqs.~\eqref{eq-1} and \eqref{eq-2} might suggest that $SU(4)$ or $SU(5)$ flavour symmetry is being assumed. In fact, only the $q\bar q$ character of the mesons involved is employed, and the vertices can be equally evaluated from the $q\bar q$ content of each meson, as shown in Ref.~\cite{Sakai:2017avl}. Moreover, as shown in Appendix~\ref{appendix}, the dominant transition matrix elements (the $-1$ off-diagonal entries in all Tables~\ref{coeff_cc_PB}--\ref{coeff_bc_PB32}) arise from the exchange of light vector mesons. The exchange of heavy vector mesons, corresponding to the terms involving the coefficients $\lambda_i$ in those tables, is suppressed by a factor $(m_V/m_H)^2$, where $m_H$ is the mass of the exchanged heavy vector meson and $m_V$ is the mass scale characterizing the interaction range, which enters through the coupling $g$ in Eqs.~\eqref{eq-1} and \eqref{eq-2} and is typically of the order of the masses of the lightest vector mesons. When light vectors are exchanged, the heavy quarks act as mere spectators in the evaluation of the matrix elements. Consequently, only the $SU(3)$ content of Eqs.~\eqref{eq-1} and \eqref{eq-2} is effectively used, and heavy-quark symmetry is preserved, since the interaction does not depend on the flavour of the heavy quark involved.

It is worth stressing that the kernel in Eq.~\eqref{eq:def_Vij} corresponds to the $s$-wave projection of a spin-independent interaction. As a consequence, all spin-parity sectors receive identical contributions, leading to degenerate spectra among the corresponding $J^P$ channels within a given meson-baryon configuration. In particular, in the $VB(\frac12^+)$ sector we have degeneracy with $J^P=\frac12^-,\frac32^-$, and in the $VB(\frac32^+)$ sector with $J^P=\frac12^-,\frac32^-,\frac52^-$.

Once the transition potential of Eq.~\eqref{eq:def_Vij} is determined, the scattering amplitude is obtained by solving the Bethe--Salpeter equation in coupled channels, which in the matrix form reads 
\begin{eqnarray}
\label{T-matrix}
   T = [1-VG]^{-1}V
\end{eqnarray}
 where $G$ is the diagonal loop function for the propagating  meson-baryon intermediate states. The loop functions are regularized with the cutoff method, as in \cite{Debastiani:2017ewu} using a cutoff ($q_{max}$) set to a value around $700$~MeV. To estimate the theoretical uncertainties and assess the stability of our results, we examine their dependence on moderate variations of the cutoff parameter. One may wonder whether the results could also be affected by other quantities, such as the couplings or decay constants entering the interaction kernel, which determine the strength of the potential. However, changes in the kernel can be compensated for, to some extent, by variations of the cutoff parameter $q_{\max}$. To illustrate this point, let us consider a single channel for simplicity. According to Eq.~\eqref{T-matrix}, a variation of $V^{-1}$ can be compensated for by a corresponding change in $G$, induced by varying $q_{\max}$, such that the scattering amplitude remains unchanged in the energy region of interest. In this sense, variations of the cutoff parameter can effectively account for changes in the strength of the interaction, providing a useful way to assess the stability of the results.

An interesting feature common to all the tables containing the $C_{ij}$ coefficients in Appendix~\ref{appendix} is that the diagonal elements identically vanish or are negligible. This implies that the channels discussed above do not interact through direct diagonal transitions and, consequently, cannot generate bound states on their own. The dynamical origin of the binding is instead provided by the coupled-channel transitions, in particular by those connected through coefficients $C_{ij}=-1$, which ultimately generate enough attraction to produce bound states, as will be shown below.
 
 This is well documented in literature. In the case of a two-channel system, one can formally eliminate one of the channels and derive an effective interaction in the remaining one \cite{Hyodo:2013nka,Wang:2022pin,Aceti:2014ala}
 \begin{equation}
 \label{V_eff}
 V_{eff}=V_{11}+\frac{V_{12}^2G_2}{1-V_{22}G_2}\,.
 \end{equation}
Because the loop function satisfies $G_2<0$ in the relevant energy region, the resulting effective potential is attractive, thereby providing the necessary mechanism for the dynamical generation of bound states.\\

\section{Results and discussion}
The aim of the present study is to search for molecular exotic pentaquarks arising from coupled-channel dynamics. Such generated states manifest themselves as pole singularities of the scattering amplitude on the second Riemann sheet at a complex energy $z_R=M_R-{\rm i}\Gamma_R/2$, whose real and imaginary parts correspond to the mass, ($M_R$), and the half width, ($\Gamma_R/2$), of the state, respectively. 

The poles are obtained on the second Riemann sheet by replacing $G\to G^{II}$ according to
\begin{eqnarray}
G^{II}_j = G^I_j + i \frac{2M_j\,q_j}{4\pi\sqrt{s}}\,,
\end{eqnarray}
for Re$\sqrt s> m_j+M_j$, and $q_j$ given by  $q_j=\lambda^{1/2}(s,m^2_j,M^2_j)/2\sqrt{s}$, with $m_j$ and $M_j$ the masses of the meson and baryon, respectively. We also evaluate the couplings, $g_i$, from the residues of the amplitudes at the pole position, where $T_{ij} = g_i g_j/(z-z_R)$. Fixing the sign of one coupling determines the relative signs of the remaining ones unambiguously. In addition, we present the quantities $g_iG^{II}_i$, which are proportional to the wave function at the origin in coordinate space~\cite{Gamermann:2009uq}.

In Tables~\ref{Table_poles_scc}, \ref{Table_poles_sbb} and \ref{Table_poles_scb} we write the pole locations, together with the couplings to each channel and
the wave function at the origin for all possible sectors classified depending on their quark content $u\bar d sQ_1Q_2$. 


\begin{table*}[t]
\caption{Pole positions (mass and half-width, in MeV) for the sector with quark content $u\bar{d}scc$, for the cutoffs $q_{\max}=650,\,700,\,750$ MeV, together with the couplings $g_i$ to the corresponding coupled channels and the associated wave functions at the origin, $g_iG^{II}_i$ (in MeV). The Riemann sheet (RS) of each pole is given in brackets, with the channels ordered by increasing threshold. A $+$ sign means that the loop function of channel $j$ is evaluated on the physical RS, $G_j = G^I_j$, while a $-$ sign means $G_j = G^I_j + i\,2M_j\,q_j/(4\pi\sqrt{s})$. Both $VB(\tfrac{1}{2}^+)$ and $VB(\tfrac{3}{2}^+)$ present virtual states for $q_{\max}=650$ MeV, which show up as bumps near the $\Omega^+_{cc}\rho^+$ and $\Omega^{*+}_{cc}\rho^+$ thresholds, respectively, once the widths of the $\rho$ and $\bar K^*$ are taken into account (see Fig.~\ref{fig3} for the $VB(\tfrac{1}{2}^+)$ sector). For these virtual states, $g_i$ and $g_iG^{II}_i$ are not listed, to discourage their use in Breit-Wigner parametrizations of the corresponding bumps.}
\label{Table_poles_scc}
\footnotesize
\renewcommand{\arraystretch}{1.1}
\setlength{\tabcolsep}{4pt}
\begin{ruledtabular}
\begin{tabular}{cccccccc}
MB sector & $q_{\max}$ & $M_R-\mathrm{i}\Gamma_R/2$\hspace{0.5em}[RS] & & \multicolumn{4}{c}{Coupled channels} \\
\hline
\hline
\multirow{7}{*}{\shortstack{$VB(\tfrac{1}{2}^+)$ \\[2pt] $J^P=\tfrac{1}{2}^-,\tfrac{3}{2}^-$}}
 & & & & $\Omega^+_{cc}\rho^+$ & $\Xi^{++}_{cc}\bar{K}^{*0}$ & $\Xi^+_c D^{*+}$ & $\Xi'^+_c D^{*+}$ \\
 & \multirow{2}{*}{650} & \multirow{2}{*}{$4484.77-\mathrm{i}0.28$\hspace{0.5em}[$+-++$]} & $g_i$ & - & - & - & - \\
 & & & $\gG$ &  - & - & - & -  \\
\cline{2-8}
 & \multirow{2}{*}{700} & \multirow{2}{*}{$4483.33-\mathrm{i}0.39$\hspace{0.5em}[$-+++$]} & $g_i$ & $-0.96+\mathrm{i}0.31$ & $1.16-\mathrm{i}0.03$ & $-0.29-\mathrm{i}0.00$ & $-0.02+\mathrm{i}0.00$ \\
 & & & $\gG$ & $23.92-\mathrm{i}0.46$ & $-19.51+\mathrm{i}0.45$ & $5.44-\mathrm{i}1.31$ & $0.15-\mathrm{i}0.00$ \\
\cline{2-8}
 & \multirow{2}{*}{750} & \multirow{2}{*}{$4477.91$\hspace{0.5em}[$++++$]} & $g_i$ & $-1.05$ & $1.22$ & $-0.30$ & $-0.02$ \\
 & & & $\gG$ & $25.12$ & $-21.50$ & $6.08$ & $0.15$ \\
\hline
\multirow{7}{*}{\shortstack{$VB(\tfrac{3}{2}^+)$ \\[2pt] $J^P=\tfrac{1}{2}^-,\tfrac{3}{2}^-,\tfrac{5}{2}^-$}}
 & & & & $\Omega^{*+}_{cc}\rho^+$ & $\Xi^{*++}_{cc}\bar{K}^{*0}$ & $\Xi^{*+}_c D^{*+}$ & \\
 & \multirow{2}{*}{650} & \multirow{2}{*}{$4541.16$\hspace{0.5em}[$-++$]} & $g_i$ &  - & - & - \\
 & & & $\gG$ & - &  - & - &  \\
\cline{2-8}
 & \multirow{2}{*}{700} & \multirow{2}{*}{$4541.02$\hspace{0.5em}[$+++$]} & $g_i$ & $-0.84$ & $1.01$ & $0.04$ & \\
 & & & $\gG$ & $21.52$ & $-17.76$ & $-0.25$ & \\
\cline{2-8}
 & \multirow{2}{*}{750} & \multirow{2}{*}{$4536.52$\hspace{0.5em}[$+++$]} & $g_i$ & $-1.21$ & $1.39$ & $0.04$ & \\
 & & & $\gG$ & $29.64$ & $-25.65$ & $-0.32$ & \\
\end{tabular}
\end{ruledtabular}
\end{table*}

\begin{table*}[t]
\caption{Same as Table~\ref{Table_poles_scc}, but for the sector with quark content $u\bar{d}sbb$.}
\label{Table_poles_sbb}
\footnotesize
\renewcommand{\arraystretch}{1.1}
\setlength{\tabcolsep}{4pt}
\begin{ruledtabular}
\begin{tabular}{cccccccc}
MB sector & $q_{\max}$ & $M_R-\mathrm{i}\Gamma_R/2$\hspace{0.5em}[RS] & & \multicolumn{4}{c}{Coupled channels} \\
\hline
\hline
\multirow{7}{*}{\shortstack{$VB(\tfrac{1}{2}^+)$ \\[2pt] $J^P=\tfrac{1}{2}^-,\tfrac{3}{2}^-$}}
 & & & & $\Omega^-_{bb}\rho^+$ & $\Xi^0_{bb}\bar{K}^{*0}$ & & \\
 & \multirow{2}{*}{650} & \multirow{2}{*}{$10999.34$\hspace{0.5em}[$++$]} & $g_i$ & $-0.71$ & $0.91$ & & \\
 & & & $\gG$ & $19.28$ & $-15.14$ & & \\
\cline{2-8}
 & \multirow{2}{*}{700} & \multirow{2}{*}{$10995.00$\hspace{0.5em}[$++$]} & $g_i$ & $-1.11$ & $1.33$ & & \\
 & & & $\gG$ & $28.36$ & $-23.71$ & & \\
\cline{2-8}
 & \multirow{2}{*}{750} & \multirow{2}{*}{$10988.17$\hspace{0.5em}[$++$]} & $g_i$ & $-1.34$ & $1.54$ & & \\
 & & & $\gG$ & $33.16$ & $-28.80$ & & \\
\hline
\multirow{7}{*}{\shortstack{$VB(\tfrac{3}{2}^+)$ \\[2pt] $J^P=\tfrac{1}{2}^-,\tfrac{3}{2}^-,\tfrac{5}{2}^-$}}
 & & & & $\Omega^{*-}_{bb}\rho^+$ & $\Xi^{*0}_{bb}\bar{K}^{*0}$ & & \\
 & \multirow{2}{*}{650} & \multirow{2}{*}{$11027.72$\hspace{0.5em}[$++$]} & $g_i$ & $-0.59$ & $0.77$ & & \\
 & & & $\gG$ & $16.32$ & $-12.48$ & & \\
\cline{2-8}
 & \multirow{2}{*}{700} & \multirow{2}{*}{$11023.88$\hspace{0.5em}[$++$]} & $g_i$ & $-1.07$ & $1.30$ & & \\
 & & & $\gG$ & $27.86$ & $-22.87$ & & \\
\cline{2-8}
 & \multirow{2}{*}{750} & \multirow{2}{*}{$11017.28$\hspace{0.5em}[$++$]} & $g_i$ & $-1.31$ & $1.54$ & & \\
 & & & $\gG$ & $33.07$ & $-28.33$ & & \\
\end{tabular}
\end{ruledtabular}
\end{table*}

\begin{table*}[t]
\caption{Same as Table~\ref{Table_poles_scc}, but for the sector with quark content $u\bar{d}scb$.}
\label{Table_poles_scb}
\footnotesize
\renewcommand{\arraystretch}{1.1}
\setlength{\tabcolsep}{4pt}
\begin{ruledtabular}
\begin{tabular}{cccccccc}
MB sector & $q_{\max}$ & $M_R-\mathrm{i}\Gamma_R/2$\hspace{0.5em}[RS] & & \multicolumn{4}{c}{Coupled channels} \\
\hline
\hline
\multirow{7}{*}{\shortstack{$VB(\tfrac{1}{2}^+)$ \\[2pt] $J^P=\tfrac{1}{2}^-,\tfrac{3}{2}^-$}}
 & & & & $\Omega^0_{bc}\rho^+$ & $\Xi^+_{bc}\bar{K}^{*0}$ & $\Xi^0_b D^{*+}$ & \\
 & \multirow{2}{*}{650} & \multirow{2}{*}{$7779.33$\hspace{0.5em}[$+++$]} & $g_i$ & $-0.89$ & $1.11$ & $0.48$ & \\
 & & & $\gG$ & $22.27$ & $-17.41$ & $-5.73$ & \\
\cline{2-8}
 & \multirow{2}{*}{700} & \multirow{2}{*}{$7773.55$\hspace{0.5em}[$+++$]} & $g_i$ & $-1.22$ & $1.44$ & $0.64$ & \\
 & & & $\gG$ & $28.89$ & $-23.97$ & $-8.03$ & \\
\cline{2-8}
 & \multirow{2}{*}{750} & \multirow{2}{*}{$7765.32$\hspace{0.5em}[$+++$]} & $g_i$ & $-1.43$ & $1.63$ & $0.74$ & \\
 & & & $\gG$ & $32.95$ & $-28.37$ & $-9.60$ & \\
\hline
\multirow{7}{*}{\shortstack{$VB'(\tfrac{1}{2}^+)$ \\[2pt] $J^P=\tfrac{1}{2}^-,\tfrac{3}{2}^-$}}
 & & & & $\Omega'^0_{bc}\rho^+$ & $\Xi'^+_{bc}\bar{K}^{*0}$ & $\Xi'^0_b D^{*+}$ & \\
 & \multirow{2}{*}{650} & \multirow{2}{*}{$7815.96$\hspace{0.5em}[$+++$]} & $g_i$ & $-0.80$ & $0.97$ & $0.06$ & \\
 & & & $\gG$ & $20.61$ & $-16.88$ & $-0.32$ & \\
\cline{2-8}
 & \multirow{2}{*}{700} & \multirow{2}{*}{$7811.46$\hspace{0.5em}[$+++$]} & $g_i$ & $-1.14$ & $1.32$ & $0.06$ & \\
 & & & $\gG$ & $28.08$ & $-24.23$ & $-0.41$ & \\
\cline{2-8}
 & \multirow{2}{*}{750} & \multirow{2}{*}{$7804.76$\hspace{0.5em}[$+++$]} & $g_i$ & $-1.36$ & $1.52$ & $0.06$ & \\
 & & & $\gG$ & $32.61$ & $-29.01$ & $-0.46$ & \\
\hline
\multirow{7}{*}{\shortstack{$VB(\tfrac{3}{2}^+)$ \\[2pt] $J^P=\tfrac{1}{2}^-,\tfrac{3}{2}^-,\tfrac{5}{2}^-$}}
 & & & & $\Omega^{*0}_{bc}\rho^+$ & $\Xi^{*+}_{bc}\bar{K}^{*0}$ & $\Xi^{*0}_b D^{*+}$ & \\
 & \multirow{2}{*}{650} & \multirow{2}{*}{$7835.83$\hspace{0.5em}[$+++$]} & $g_i$ & $-0.53$ & $0.68$ & $0.05$ & \\
 & & & $\gG$ & $14.56$ & $-11.28$ & $0.07$ & \\
\cline{2-8}
 & \multirow{2}{*}{700} & \multirow{2}{*}{$7832.31$\hspace{0.5em}[$+++$]} & $g_i$ & $-1.06$ & $1.27$ & $0.07$ & \\
 & & & $\gG$ & $27.13$ & $-22.50$ & $-0.47$ & \\
\cline{2-8}
 & \multirow{2}{*}{750} & \multirow{2}{*}{$7826.03$\hspace{0.5em}[$+++$]} & $g_i$ & $-1.13$ & $1.51$ & $0.07$ & \\
 & & & $\gG$ & $32.46$ & $-28.04$ & $-0.54$ & \\
\end{tabular}
\end{ruledtabular}
\end{table*}

The most salient feature emerging from Tables~\ref{Table_poles_scc}, \ref{Table_poles_sbb} and \ref{Table_poles_scb} is the clear pattern exhibited by the pole positions and couplings. The dynamically generated molecular states systematically appear below the $\Xi^{(')(*)}_{Q_1Q_2}\bar{K}^{(*)0}$ thresholds, in most cases lying below the lowest threshold, while in few others being located between nearby channels, thus manifesting themselves either as bound states or resonances depending on the sector considered. A characteristic feature is the sizable coupling of these states to the $\Omega^{(')(*)}_{Q_1Q_2}\pi^+(\rho^+)$ and $\Xi^{(')(*)}_{Q_1Q_2}\bar{K}^{(*)0}$ channels, with the latter generally displaying slightly larger couplings. The dominant role played by these channels in the molecular structure is further supported by the corresponding values of the wave function at the origin, whose magnitudes are significantly larger than those associated with the remaining coupled channels, whenever present.

These findings are naturally understood from the structure of the coefficients collected in Appendix~\ref{appendix}. In particular, the transition coefficients ($C_{\Omega^{(')(*)}_{Q_1Q_2}\pi^+(\rho^+),\Xi^{(')(*)}_{Q_1Q_2}\bar{K}^{(*)0}}\coloneqq C_{12}$) are the largest ones, thereby enhancing the corresponding off-diagonal interaction kernel, Eq.~\eqref{eq:def_Vij}, denoted here for convenience as $V_{12}$. This contribution enters directly into the effective potential of Eq.~\eqref{V_eff}. As discussed above, and restricting the discussion for simplicity to an effective two-channel system --- a reasonable approximation given the subleading role played by the remaining transitions --- the characteristic situation in which the diagonal terms vanish, $V_{11}=V_{22}=0$, while the transition potential $V_{12}$ is sufficiently large, naturally leads to enough attraction to dynamically generate bound states or resonances. Indeed, according to Eq.~\eqref{V_eff}, the condition $G_2<0$ for the loop function of the heavier channel guarantees that the induced effective interaction becomes attractive.

An additional observation concerns the strong suppression of the channels involving additional heavy mesons, such as $\Xi^{(')(*)}_{c}D^{(*)}$ and $\Xi^{(')(*)}_{b}D^{(*)}$, whose couplings and wave functions at the origin are systematically smaller than those associated with the $\Omega^{(')(*)}_{Q_1Q_2}\pi^+(\rho^+)$ and $\Xi^{(')(*)}_{Q_1Q_2}\bar K^{(*)0}$ channels. This hierarchy further supports the interpretation of the generated states as molecular configurations driven predominantly by the interplay of the latter two channels. Moreover, the results display a highly systematic pattern across the $u\bar d scc$, $u\bar d scb$, and $u\bar d sbb$ sectors, suggesting that the same coupled-channel mechanism is responsible for the dynamical generation of all these states independently of the heavy-quark content.

We estimate the stability and uncertainties in our results arising from the dependence on the cutoff parameter. The maximum momentum in the loop integral, $q_{\text{max}}$, has been fixed to $700$~MeV, as justified above, which provides a reasonable starting point for a realistic description of the exotic states generated within this coupled-channel framework. To assess the associated uncertainties and check the stability, we repeat the calculations varying the cutoff in the range $q_{\text{max}}=650$--$750$~MeV, as can be appreciated in Tables~\ref{Table_poles_scc}, \ref{Table_poles_sbb} and \ref{Table_poles_scb}. In this analysis, the dependence of the pole positions and coupling constants on the cutoff parameter $q_{\max}$ has also been examined. For the $u\bar{d}scc$ sector, we exclude the results obtained with $q_{\max}=650$~MeV from this comparison, since they correspond to virtual states. The pole masses exhibit a rather weak cutoff dependence in all sectors, with maximum variations ranging from approximately $5$ to $14$~MeV. In particular, the largest mass variation is found in the $VB(1/2^+)$ sector with $u\bar{d}scb$ quark content. The coupling constants show a somewhat stronger dependence on the cutoff. In the $u\bar{d}scc$ sector, the maximum relative variations are approximately $9\%$ and $44\%$ in the $VB(1/2^+)$ and $VB(3/2^+)$ sectors, respectively. Larger variations are observed in the $u\bar{d}sbb$ and $u\bar{d}scb$ sectors, reaching approximately $122\%$ in the \(VB(3/2^+)\) sectors. Overall, the pole masses remain remarkably stable against variations of the cutoff, whereas the coupling constants exhibit a more noticeable, although channel-dependent, sensitivity to the regularization parameter.

In the sectors not listed in Tables~\ref{Table_poles_scc}, \ref{Table_poles_sbb} and \ref{Table_poles_scb}, the amplitudes display clear cusp structures at the thresholds of the channels that open in the energy region considered. The cusp itself is a general consequence of unitarity: the amplitude has a square-root branch point at each S-wave threshold (Wigner cusp). However, whether it shows up as a pronounced structure in the line shape depends on the strength of the attraction in the coupled-channel dynamics. In this sense, within the molecular picture, these signals are not purely kinematical effects but reflect the underlying coupled-channel interaction. Comparable structures are well known in unitarized EFT approaches. A paradigmatic example is the $a_0(980)$, which in Refs.~\cite{Oller:1997ti,Baru:2003qq} is described as a $\pi\eta$-$K\bar K$ molecular state with a pronounced cusp at the $K\bar K$ threshold. In the present work, analogous structures appear at different thresholds depending on the sector. These thresholds, together with the associated virtual-state pole positions, are listed in Table~\ref{Virtual_states}.
\begin{table*}[t]
\caption{Virtual states found in the $PB$ sectors with $u\bar{d}scc$, $u\bar{d}sbb$ and $u\bar{d}scb$ quark content, for the momentum cutoffs $q_{\max}=650,\,700,\,750$ MeV. For each state we give the pole position $M_R-\mathrm{i}\Gamma_R/2$ (in MeV) and, in brackets, the Riemann sheet (RS) on which it is found, following the convention of Table~\ref{Table_poles_scc}. The last column gives the threshold at which the state shows up as a cusp-like structure in the amplitudes.}
\label{Virtual_states}
\footnotesize
\renewcommand{\arraystretch}{1.5}
\setlength{\tabcolsep}{6pt}
\begin{ruledtabular}
\begin{tabular}{cccccc}
MB sector & $J^P$ & Cusp at threshold
 & $q_{\max}=650$ & $q_{\max}=700$ & $q_{\max}=750$ \\
\hline
\multicolumn{6}{c}{$u\bar{d}scc$} \\
\hline
$PB(\tfrac{1}{2}^+)$ & $\tfrac{1}{2}^-$ & $\Xi^{++}_{cc}\bar{K}^0$
 & $3979.01-\mathrm{i}143.15$ [$--++$]
 & $3979.92-\mathrm{i}134.45$ [$--++$]
 & $3979.98-\mathrm{i}125.58$ [$--++$] \\
$PB(\tfrac{3}{2}^+)$ & $\tfrac{3}{2}^-$ & $\Xi^{*++}_{cc}\bar{K}^0$
 & $4035.44-\mathrm{i}144.30$ [$--+$]
 & $4036.19-\mathrm{i}135.92$ [$--+$]
 & $4036.54-\mathrm{i}127.38$ [$--+$] \\
\hline
\multicolumn{6}{c}{$u\bar{d}sbb$} \\
\hline
$PB(\tfrac{1}{2}^+)$ & $\tfrac{1}{2}^-$ & $\Xi^0_{bb}\bar{K}^0$
 & $10488.10-\mathrm{i}126.05$ [$--$]
 & $10488.32-\mathrm{i}117.86$ [$--$]
 & $10487.74-\mathrm{i}109.62$ [$--$] \\
$PB(\tfrac{3}{2}^+)$ & $\tfrac{3}{2}^-$ & $\Xi^{*0}_{bb}\bar{K}^0$
 & $10517.13-\mathrm{i}125.94$ [$--$]
 & $10517.35-\mathrm{i}117.76$ [$--$]
 & $10516.76-\mathrm{i}109.54$ [$--$] \\
\hline
\multicolumn{6}{c}{$u\bar{d}scb$} \\
\hline
$PB(\tfrac{1}{2}^+)$ & $\tfrac{1}{2}^-$ & $\Xi^+_{bc}\bar{K}^0$
 & $7270.11-\mathrm{i}127.06$ [$--+$]
 & $7269.86-\mathrm{i}117.99$ [$--+$]
 & $7268.63-\mathrm{i}108.83$ [$--+$] \\
$PB'(\tfrac{1}{2}^+)$ & $\tfrac{1}{2}^-$ & $\Xi'^+_{bc}\bar{K}^0$
 & $7304.50-\mathrm{i}130.84$ [$--+$]
 & $7304.99-\mathrm{i}122.54$ [$--+$]
 & $7304.66-\mathrm{i}114.16$ [$--+$] \\
$PB(\tfrac{3}{2}^+)$ & $\tfrac{3}{2}^-$ & $\Xi^{*+}_{bc}\bar{K}^0$
 & $7325.60-\mathrm{i}130.59$ [$--+$]
 & $7326.06-\mathrm{i}122.32$ [$--+$]
 & $7325.71-\mathrm{i}113.95$ [$--+$] \\
\end{tabular}
\end{ruledtabular}
\end{table*} 

It is also worth noting that the states generated in the \(VB(\tfrac{1}{2}^+)\) and \(VB(\tfrac{3}{2}^+)\) sectors with \(u\bar{d}scc\) quark content, which couple predominantly to the \(\Omega^+_{cc}\rho^+\)–\(\Xi^{++}_{cc}\bar{K}^{*0}\) and \(\Omega^{*+}_{cc}\rho^+\)–\(\Xi^{*++}_{cc}\bar{K}^{*0}\) channels, respectively, are compatible with some of the states reported in Ref.~\cite{Yang:2024okq}. More specifically, the states reported in that reference with \(J^P=\tfrac{3}{2}^-\), originating from the \(\Omega^+_{cc}\rho^+\) or \(\Xi^{++}_{cc}\bar{K}^{*0}\) configurations, are compatible with our \(VB(\tfrac{1}{2}^+)\) results, while those with \(J^P=\tfrac{5}{2}^-\), originating from the \(\Omega^{*+}_{cc}\rho^+\) or \(\Xi^{*++}_{cc}\bar{K}^{*0}\) configurations, are compatible with our \(VB(\tfrac{3}{2}^+)\) results. We note that the two approaches are conceptually different and complementary: our work is based on a coupled-channel molecular picture in which the states appear as dynamically generated poles of the scattering amplitude, whereas Ref.~\cite{Yang:2024okq} employs QCD sum rules and also considers compact configurations. The agreement found between both approaches for some of the states therefore provides additional support for their existence.

\begin{figure}   
  \centering
  \includegraphics[width=0.5\textwidth]{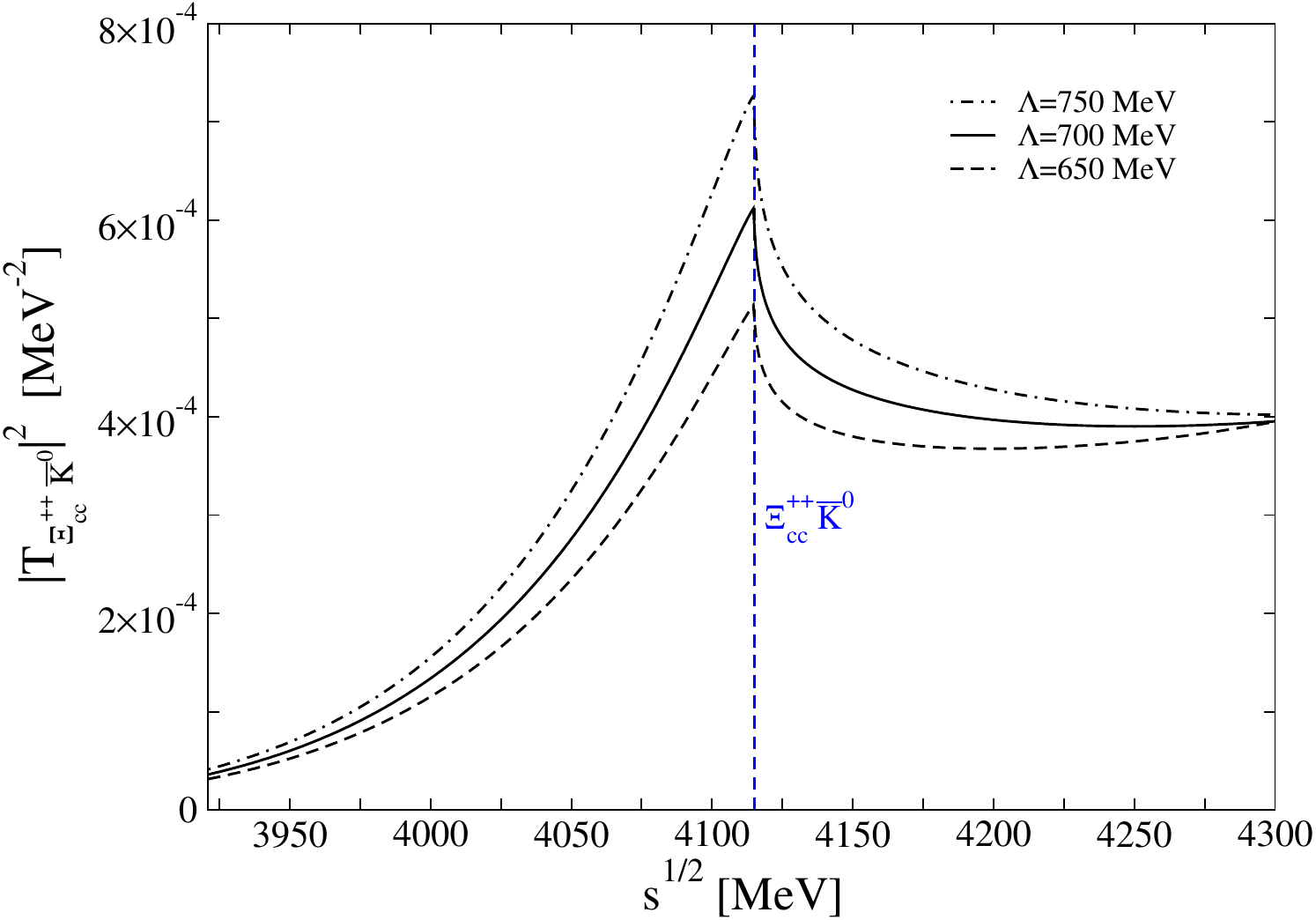}
\caption{Squared absolute value of the elastic $\Xi^{++}_{cc}\bar{K}^0 \to \Xi^{++}_{cc}\bar{K}^0$ scattering amplitude for $q_{max}=650$, $700$ and $750$~MeV. The vertical lines indicate the thresholds of the channels included in the calculation that lie within the energy range shown.}
  \label{fig2}
\vspace{-0.5cm}
\end{figure}

\begin{figure}   
  \centering
  \includegraphics[width=0.5\textwidth]{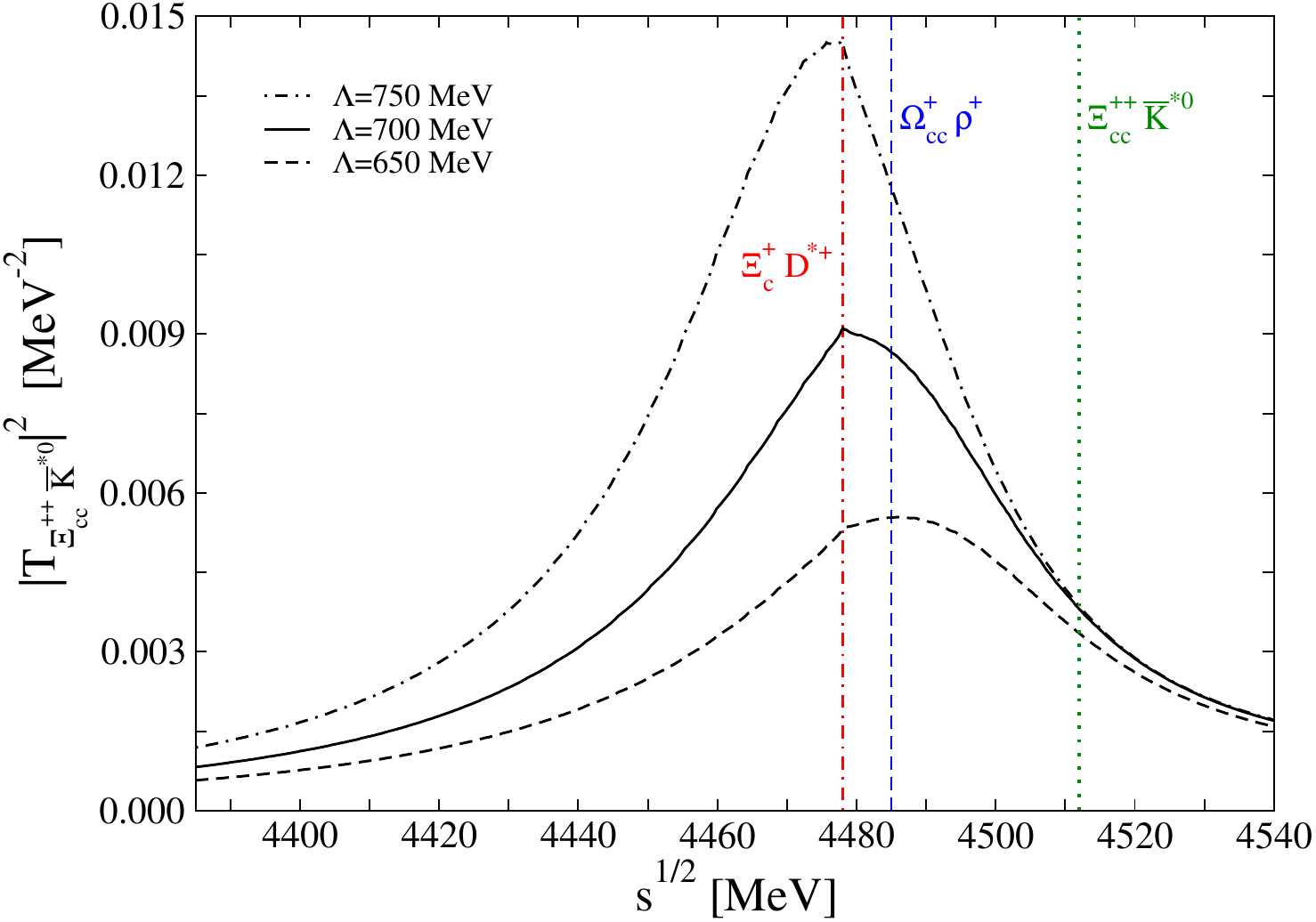}
 \caption{Squared absolute value of the elastic $\Xi^{++}_{cc}\bar{K}^{*0} \to \Xi^{++}_{cc}\bar{K}^{*0}$ scattering amplitude for $q_{max}=650$, $700$ and $750$~MeV. The vertical lines indicate the thresholds of the channels included in the calculation that lie within the energy range shown.}
  \label{fig3}
\vspace{-0.5cm}
\end{figure}

In order to complement the analysis of the pole content, we present the squared absolute value of two scattering amplitudes to illustrate how the signals clearly manifest themselves both in the presence of poles and in the case of cusp structures. For this purpose, we have selected the elastic $\Xi^{++}_{cc}\bar{K}^0$ and $\Xi^{++}_{cc}\bar{K}^{*0}$ channels, belonging respectively to the $PB(\frac{1}{2}^+)$ and $VB(\frac{1}{2}^+)$ sectors with $u\bar d scc$ quark content, shown in Figs.~\ref{fig2} and \ref{fig3}, respectively. To assess the stability of these structures against variations of the regularization parameter, we also display the squared moduli of the amplitudes for $q_{max}=650$, $700$ and $750$~MeV in both figures. It should be noted that, to obtain more realistic amplitudes, we incorporate the finite widths of the $\bar{K}^{*0}$ and $\rho^+$ mesons in the channels involving them by folding the loop functions with the corresponding spectral functions, following the procedure described in Eqs.~(7)--(9) of Ref.~\cite{Lin:2025mtz}.

\section{Summary}
Altogether, the present formalism predicts two molecular pentaquarks in the $u\bar d scc$ sector, two such states in the $u\bar d sbb$ sector, and three in the $u\bar d scb$ sector. Owing to the spin-independent nature of the interaction kernel, several of these states appear as degenerate spin multiplets. In particular, the states generated from the $VB(\frac{1}{2}^+)$ sectors are degenerate with $J^P=\frac{1}{2}^-,\frac{3}{2}^-$, while those originating from the $VB(\frac{3}{2}^+)$ sectors are degenerate with $J^P=\frac{1}{2}^-,\frac{3}{2}^-,\frac{5}{2}^-$. \\

Regarding the experimental strategies for observing these states, we suggest searching for them through the invariant-mass distributions of two-particle channels whose thresholds lie closest to the predicted state masses and to which the states couple strongly. Such measurements could be performed by the LHCb Collaboration in high-energy, high-multiplicity \(pp\) collisions, where invariant-mass distributions of many analogous two-particle systems have already been measured, as reported in Refs.~\cite{LHCb:2015yax,LHCb:2017uwr}.

\section{Acknowledgment}
This work is supported by the Spanish Ministerio de Ciencia e Innovaci\'on (MICINN) under contracts PID2020-112777GB-I00, PID2023-147458NB-C21 and CEX2023-001292-S; by Generalitat Valenciana under contracts PROMETEO/2020/023 and  CIPROM/2023/59. The authors thank the warm support of the ACVJLI.

\appendix
\section{ $C_{ij}$ coefficients}
\label{appendix}

The following tables contain the $C_{ij}$ coefficients of the transition potential in Eq.~\eqref{eq:def_Vij} among the coupled channels.
\vspace{-0.4cm}
\begin{table}[H]
\centering
 \caption{Coefficients $C_{ij}$ for the $PB(\frac{1}{2}^+)$ sector of $u\bar d scc$ states.}
 \label{coeff_cc_PB}
\setlength{\tabcolsep}{6.5pt}
\begin{tabular}{l|cccc}
\hline
\hline  
  & $\Omega^+_{cc}\pi^+$ & $\Xi^{++}_{cc}\bar{K}^0$  & $\Xi^+_c D^+$ & $\Xi'^+_c D^+$ \\
\hline
 $\Omega^+_{cc}\pi^+$       & $0$   & $-1$   & $\frac{\sqrt{3}}{2\sqrt{2}}\lambda_1$  & $\frac{1}{2\sqrt{2}}\lambda_1$   \\
 $\Xi^{++}_{cc}\bar{K}^0$  &          & $0$                                       & $\frac{-\sqrt{3}}{2\sqrt{2}}\lambda_2$                     & $\frac{1}{2\sqrt{2}}\lambda_2$  \\
  $\Xi^+_c D^+$                   &          &                                              & $-\lambda_3$                                                    & 0                    \\
  $\Xi'^+_c D^+$                  &          &                                              &                                                            & $-\lambda_3$           \\
\hline
\hline
\end{tabular}
\end{table}
The coefficients $\lambda_1$, $\lambda_2$, and $\lambda_3$ account for the suppression factors associated with the exchange of $D^*$, $D_s^*$, or $J/\psi$ mesons, respectively, instead of a $K^*$ meson, as in the case of $\Omega_{cc}^+\pi^+ \to \Xi_{cc}^{++}\bar{K}^0$ transition. They are defined as $\lambda_1=m^2_V/m^2_{D^*}$, $\lambda_2=m^2_V/m^2_{D_s^*}$ and $\lambda_3=m^2_V/m^2_{J/\psi}$.

The matrix of $C_{ij}$ coefficients for the case of  $VB(\frac{1}{2}^+)$ sector is the same as Table~\ref{coeff_cc_PB}, replacing $\pi^+\to\rho^+$, $\bar K^0\to \bar K^{*0}$ and $D^+ \to D^{*+}$.
\vspace{-0.4cm}
\begin{table}[H]
\centering
 \caption{Coefficients $C_{ij}$ for the $PB(\frac{3}{2}^+)$ sector of $u\bar d scc$ states.}
 \label{coeff_cc_PB32}
\setlength{\tabcolsep}{6.5pt}
\begin{tabular}{l|ccc}
\hline
\hline  
  & $\Omega^{*+}_{cc}\pi^+$ & $\Xi^{*++}_{cc}\bar{K}^0$ & $\Xi^{*+}_c D^+$  \\
\hline
$\Omega^{*+}_{cc}\pi^+$       & $0$   & $-1$   & $\frac{-1}{\sqrt{2}}\lambda_1$   \\
 $\Xi^{*++}_{cc}\bar{K}^0$  &          & $0$  & $\frac{-1}{\sqrt{2}}\lambda_2$  \\
 $\Xi^{*+}_c D^+$           &       &   & $-\lambda_3$                    \\
\hline
\end{tabular}
\end{table}
The $C_{ij}$ coefficients required for the $VB(\frac{3}{2}^+)$ sector are identical to those collected in Table~\ref{coeff_cc_PB32}. The corresponding channels are obtained from those listed in Table~\ref{coeff_cc_PB32} by performing the replacements $\pi^+\to\rho^+$, $\bar K^0\to \bar K^{*0}$, and $D^+\to D^{*+}$.

The case with $u\bar d s b b$ quark content can be treated analogously. However, the exchange of a $D^*$ meson is now replaced by $B^*$ exchange, while $D_s^*$ exchange becomes $B_s^*$ exchange. Likewise, the already strongly suppressed $J/\psi$ exchange is replaced by $\Upsilon$ exchange, which is entirely negligible. Owing to the large masses of these exchanged vector mesons, the corresponding transition amplitudes are highly suppressed. Consequently, we neglect the transitions mediated by $B^*$, $B_s^*$, and $\Upsilon$ mesons. As a result, the $\Xi_b^0\bar B^0$ and $\Xi_b^{'0}\bar B^0$ channels decouple both from the remaining coupled channels and from each other. Therefore, the dynamics is effectively driven only by the $\Omega_{bb}^-\pi^+$ and $\Xi_{bb}^0\bar K^0$ channels, whose corresponding coupling coefficients are shown in Table~\ref{coeff_bb_PB}.
\vspace{-0.4cm}
\begin{table}[H]
\centering
 \caption{Coefficients $C_{ij}$ for the $PB({\frac{1}{2}}^+)$ sector of $u\bar d sbb$ states.}
 \label{coeff_bb_PB}
 \setlength{\tabcolsep}{6.5pt}
\begin{tabular}{l|cc}
\hline
\hline
 ~  & $\Omega^-_{bb}\pi^+$ & $\Xi^0_{bb}\bar{K}^0$ \\
\hline
  $\Omega^-_{bb}\pi^+$     &  $0$    & $-1$        \\
  $\Xi^0_{bb}\bar{K}^0$            &             &$0$        \\
\hline
\hline
\end{tabular}
\end{table}
Once again, the same matrix applies to the $VB(\frac{1}{2}^+)$ sector, with the substitutions $\pi^+\to\rho^+$ and $\bar K^0\to \bar K^{*0}$.

For the $VB(\frac{3}{2}^+)$ sector the situation is analogous, and one obtains the same $C_{ij}$ coefficients as those shown in Table~\ref{coeff_bb_PB32}.
\vspace{-0.4cm}
\begin{table}[H]
\centering
 \caption{Coefficients $C_{ij}$ for the $PB({\frac{3}{2}}^+)$ sector of $u\bar d sbb$ states.}
 \label{coeff_bb_PB32}
 \setlength{\tabcolsep}{6.5pt}
\begin{tabular}{l|cc}
\hline
\hline
 ~  & $\Omega^{*-}_{bb}\pi^+$ & $\Xi^{*0}_{bb}\bar{K}^0$ \\
\hline
  $\Omega^{*-}_{bb}\pi^+$     &  $0$    & $-1$        \\
  $\Xi^{*0}_{bb}\bar{K}^0$            &             &$0$        \\
\hline
\hline
\end{tabular}
\end{table}
Finally, in the $u\bar d s b c$ sector, the only channels that can be removed from the coupled-channel basis are $\Xi_c^+ \bar{B}^0$ in the $PB(\frac{1}{2}^+)$ sector and $\Xi_c^{\prime +} \bar{B}^0$ in the $PB'(\frac{1}{2}^+)$ sector of Table~\ref{Table_QQs}, since they play no relevant role in the dynamics. Indeed, any transition involving these channels requires the exchange of bottomed vector mesons and is therefore strongly suppressed. The same situation also applies to the remaining fourth channels listed in Table~\ref{Table_QQs} for the $u\bar d s c b$ sector. The corresponding $C_{ij}$ couplings are collected in Tables~\ref{coeff_bc_PB} and~\ref{coeff_bc_PB'}.
\vspace{-0.4cm}
\begin{table}[H]
\centering
 \caption{Coefficients $C_{ij}$ for the $PB(\frac{1}{2}^+)$ sector of $u\bar d scb$ states.}
 \label{coeff_bc_PB}
\setlength{\tabcolsep}{6.5pt}
\begin{tabular}{l|ccc}
\hline
\hline  
  & $\Omega^0_{bc}\pi^+$  & $\Xi^+_{bc}\bar{K}^0$   & $\Xi^0_b D^+$ \\
\hline
 $\Omega^0_{bc}\pi^+$   & $0$   & $-1$   & $-\lambda_1$    \\
 $\Xi^+_{bc}\bar{K}^0$  &          & $0$    & $\lambda_2$   \\
  $\Xi^0_b D^+$         &          &        & $0$           \\
\hline
\hline
\end{tabular}
\end{table}
\vspace{-0.4cm}
\begin{table}[H]
\centering
 \caption{Coefficients $C_{ij}$ for the $PB'(\frac{1}{2}^+)$ sector of $u\bar d scb$ states.}
 \label{coeff_bc_PB'}
\setlength{\tabcolsep}{6.5pt}
\begin{tabular}{l|ccc}
\hline
\hline  
  & $\Omega'^0_{bc}\pi^+$  & $\Xi'^+_{bc}\bar{K}^0$   & $\Xi'^0_b D^+$ \\
\hline
 $\Omega'^0_{bc}\pi^+$   & $0$   & $-1$   & $-\lambda_1$    \\
 $\Xi'^+_{bc}\bar{K}^0$  &          & $0$    & $-\lambda_2$   \\
  $\Xi'^0_b D^+$         &          &        & $0$           \\
\hline
\hline
\end{tabular}
\end{table}
The change of sign of $\lambda_2$ in Table~\ref{coeff_bc_PB'} compared to that of Table~\ref{coeff_bc_PB} is due to the sandwich of the corresponding flavor wave functions, i. e. $ \langle us-su| su-us \rangle = -2$ while $ \langle us+su| su+us \rangle = 2$. 

Again, for the $VB(\frac{1}{2}^+)$ and $VB'(\frac{1}{2}^+)$ sectors, the coupling matrices remain unchanged, with the corresponding replacements in the meson channels given by $\pi^+\to\rho^+$, $\bar K^0\to \bar K^{*0}$, and $D^+\to D^{*+}$. 

\vspace{-0.4cm}
\begin{table}[H]
\centering
 \caption{Coefficients $C_{ij}$ for the $PB(\frac{3}{2}^+)$ sector of $u\bar d scb$ states.}
 \label{coeff_bc_PB32}
\setlength{\tabcolsep}{6.5pt}
\begin{tabular}{l|ccc}
\hline
\hline  
  & $\Omega^{*0}_{bc}\pi^+$  & $\Xi^{*+}_{bc}\bar{K}^0$   & $\Xi^{*0}_b D^+$ \\
\hline
 $\Omega^{*0}_{bc}\pi^+$   & $0$   & $-1$   & $-\lambda_1$    \\
 $\Xi^{*+}_{bc}\bar{K}^0$  &          & $0$    & $-\lambda_2$   \\
  $\Xi^{*0}_b D^+$         &          &        & $0$           \\
\hline
\hline
\end{tabular}
\end{table}
To conclude, for the sectors involving decuplet baryons, where one also has a symmetric flavor wave function and spin totally symmetric $\chi_S$, the matrix obtained is like in the case of Table~\ref{coeff_bc_PB'}, which we show in  Table~\ref{coeff_bc_PB32}.

As in the previous cases, the vector-baryon decuplet sector is described by the same $C_{ij}$ matrix as that given in Table~\ref{coeff_bc_PB32}, where the meson in each channel is replaced according to $\pi^+\to\rho^+$, $\bar K^0\to \bar K^{*0}$, and $D^+\to D^{*+}$.

\bibliography{refs}

\end{document}